\documentclass[sigconf]{acmart}

\usepackage{booktabs} 
\usepackage{enumitem}
\usepackage{amssymb}
\usepackage{algorithm}
\usepackage[noend]{algpseudocode}
\usepackage{subcaption}
\usepackage{soul}

\begin{document}
\vspace{-2em}
\title{Noise- and Outlier-Resistant Tomographic Reconstruction under Unknown Viewing Parameters}

\author{Ritwick Chaudhry	}
\affiliation{%
  \institution{Adobe Research}
}
\email{rchaudhr@adobe.com}

\author{Arunabh Ghosh	}
\affiliation{%
  \institution{IIT Bombay}
}
\email{arunabhghosh@iitb.ac.in}

\author{Ajit Rajwade}
\affiliation{%
  \institution{IIT Bombay}
}
\email{ajitvr@cse.iitb.ac.in}


\renewcommand{\shorttitle}{}

\begin{abstract}
In this paper, we present an algorithm for effectively reconstructing an object from a set of its tomographic projections without any knowledge of the viewing directions or any prior structural information, in the presence of pathological amounts of noise, unknown shifts in the projections, and outliers among the projections. The outliers are mainly in the form of  a number of projections of a completely different object, as compared to the object of interest. We introduce a novel approach of first processing the projections, then obtaining an initial estimate for the orientations and the shifts, and then define a refinement procedure to obtain the final reconstruction. Even in the presence of high noise variance (up to $50\%$ of the average value of the (noiseless) projections) and presence of outliers, we are able to successfully reconstruct the object. We also provide interesting empirical comparisons of our method with the sparsity based optimization procedures that have been used earlier for image reconstruction tasks.
\end{abstract}



\settopmatter{printacmref=false}
\renewcommand\footnotetextcopyrightpermission[1]{} 
\pagestyle{plain}

\maketitle

\section{Introduction}
\label{sec:intro}
Reconstructing the structure of an object from its tomographic projections (hereafter referred to as `projections') is a fundamental research problem that arises in diverse fields, such as medical imaging \cite{Fessler2000} and reconstruction in cryo-electron-microscopy  (referred to hereafter as `Cryo-EM') \cite{frank2006three}. If the viewing orientations are known \textit{a priori}, standard algorithms such as Filtered Backprojection (FBP) or its variants adapted for different acquisition geometries \cite{Feldkamp1984PracticalAlgorithm}, can be used to reconstruct the image. However, there are many scenarios where the viewing orientations are unknown. One such example is Cryo-EM where the objective is to determine the structure of a macro-molecule or a biological specimen such as a ribosome or virus, from its projections which essentially appear in various unknown orientations \cite{frank2006three,Sigworth2016PrinciplesProcessing}. Other examples include insect tomography \cite{Genise1995ApplicationTraces}, or tomography of objects performing unknown rigid motion \cite{Yu2007DataCT}. Both these are equivalent to performing  tomographic reconstruction on a fixed object, with the viewing directions being unknown. Uncertainty in viewing angles may also occur due to patient motion in medical imaging, even though to a lower degree. 
In recent times, significant research has emerged in the field of `tomography under unknown viewing parameters'. Much of this research belongs to one of the following two categories: 
\begin{enumerate}
\item Machine learning based approaches, which are based on assumptions on the distribution of the unknown parameters, typically the uniform distribution. This essentially requires that the number of projections is large. However, certain orientations might be more likely for a given structure \cite{Sigworth2016PrinciplesProcessing}, and thus would yield the aforementioned assumption unfounded. In fact, the orientation distribution may not even be known in advance. For example, in \cite{Basu2000FeasibilityAngles} a nearest neighbour search is performed over the projections to find an ordering of the angles, and then these ordered projections are assigned to angles chosen at uniform intervals on the unit half-circle (i.e. from 0 to 180 degrees). Similarly in \cite{Fang2011SLLE:Tomography}, a variant of the popular Locally Linear Embedding (LLE) algorithm, termed `Spherical LLE', is used to embed the projections on a 2D circle and finally the 2D projections are sorted and arranged in a uniform manner around the circle. A graph Laplacian based clustering approach with similar assumptions is used in \cite{Coifman2008}.

\item The other major class of algorithms have used the Moment based approach \cite{Natterer2001TheTomography} which uses the Helgason Ludwig Consistency Conditions (HLCC). These conditions relate the geometric moments of the underlying image $z(x, y)$ and the geometric moments of its tomographic projections at any angle. An alternating optimization produces estimates of the unknown angles, as well as the image moments as a by-product \cite{Malhotra2016TomographicRelationships}.
\end{enumerate}
\noindent
It has even been proved that in the case where the projections are noiseless, a unique solution exists for the given HLCC \cite{Basu2000UniquenessAngles} under some weak assumptions. However, if we apply this algorithm in practical applications it may yield very poor results. This is because practical applications often bring with them a whole new set of challenges, such as those enlisted below. 
\begin{enumerate}
\item In Cryo-EM, most biological specimens are extremely radio-sensitive, so they must be imaged with low-dose electron beams which leads to extremely high amounts of noise in the projections. The geometric moments are known to be highly sensitive to noise which compromises the entire procedure. 
\item Moreover, samples of the same biological specimen (eg: virus, ribosome, etc.) that are acquired on a single slide may often not be exactly identical due to contaminants such as ice particles as well as genuine conformational changes, and therefore there is an added complexity due to outliers \cite{Huang2016}. 
\item Some of the projections may even be shifted by a small random amount, which if not corrected, produces a very poor fidelity reconstruction. In medical tomography or insect tomography, this occurs due to subject motion. \item In Cryo-EM, the reason for the shifts is different. Here, a single slide contains potentially thousands of samples of the same biological specimen. The acquired image (often called a `micrograph') thus consists of thousands of noisy projections against a very noisy background. This necessitates a pre-processing step where the projections of the actual specimen (termed `particles') are detected via a procedure called `particle picking' \cite{Sigworth2016PrinciplesProcessing}. Since such a detection procedure will usually not be perfect, there are bound to be small translational errors between the actual and predicted particle location. This leads to uncertainty in the positions of the particles.
\end{enumerate}
\noindent
Summarily, there are four main challenges that need to be addressed, in order to accurately estimate the underlying structure from the given set of projections: (1) Severe levels of noise in the projections, (2) Unknown orientations of the projections, (3) Unknown or inaccurately known (albeit small) shifts in the projections, and (4) Presence of outliers. \\
\noindent
In this paper, we present an algorithm which (1) systematically removes the outliers, (2) clusters similar projections together, (3) uses a moments based approach to obtain an initial estimate for the orientations and the shifts, and (4) finally optimizes for the structure of the unknown object along with a refinement of the viewing parameters (i.e., the angles and the shifts). As we will see, this algorithm can successfully determine the structure of the object from its projections taken in unknown orientations despite the presence of noise, outliers and random unknown shifts. In this paper, we work exclusively with 2D images (and hence simulated 1D parallel-beam projections) for reconstruction, even though actual objects are 3D. This follows previous work in the image processing community which has studied the 2D variant of this problem extensively \cite{Basu2000FeasibilityAngles,Singer2013Two-DimensionalDirections}. Nonetheless, the underlying principles remain the same, and the computational problem remains very challenging even for reconstructing 2D images.

\section{Algorithm Description}
\label{sec:format}
\subsection{Robust Clustering}
\label{subsec:robust_clustering}
Typically in the case of Cryo-EM, a large number of projections are available. But all of these projections have pathological amounts of noise and there may even be a fair number of completely erroneous projections which owe themselves, not to the actual particle, but to the presence of foreign objects and ice particles \cite{Huang2016}. We henceforth refer to these completely erroneous projections as `outliers of \textit{Class 1}'. Reconstructing the object from these projections directly understandably yields very poor results. To combat this, we seek to cluster the projections into a small number of classes, based on orientation and structural similarity. The aim is to produce a representative set of projections that will be significantly less noisy, while simultaneously detecting and rejecting outliers of Class 1. We use the K-means  algorithm \cite{Lloyd1982LeastPCM} to cluster a large number of projections into a much smaller number of clusters $K_c$. The distance metric used is the $\ell_q$ quasi-norm ($0 < q \leq 1$) and therefore the cluster centroid is expected to  be robust to outliers. The objective function that is minimized is as follows:
\begin{equation}
(L_{centroid})(\{\xi_j\}_{j=1}^{K_c}) = \sum_{j=1}^{K_c}\sum_{z_i \in \pi_j} \|z_i - \xi_j\|_q,
\end{equation}
where there are $K_c$ clusters, $\pi_j$ represents the $j^{\textrm{th}}$ cluster and $\xi_j$ represents the $j^{\textrm{th}}$ cluster centroid. In the case when $q = 1$, the cluster centroid would be the element-wise median of the points belonging to the cluster. 

\subsection{Removal of Class 1 Outliers}
\label{subsec:class1}
After clustering, we remove some $f\%$ of the projections based on their $\ell_2$ distance from the closest cluster centroid. It is likely that since a completely erroneous projection is located far away from the other projections, a cluster will not be formed close to it. Therefore, removing $f\%$ of projections that are placed farthest from any cluster centroid will remove the Class 1 outliers. A reasonable estimate of $f$ can be provided by a biologist upon eye-balling of the micrograph, and usually, a moderate over-estimate of $f$ is not a problem.

\subsection{Averaging to form a single cluster}
\label{subsec:averaging}
Within each cluster, we define the processed projection image $\tilde{p}_j$ (for cluster index $j$) to be the average of all the projections assigned to that cluster, and which were not discarded by the previous step. This is mathematically represented as follows:
\begin{equation}
\tilde{p}_j = \dfrac{\sum_{z_i \in \pi_j} z_i(1-I_j(z_i))}{\sum_{z_i \in \pi_j} (1-I_j(z_i))}
\end{equation}
where $I_j(z_i) = 1$, when the $i^{\textrm{th}}$ projection image belonging to the $j^{\textrm{th}}$ cluster is discarded, and 0 otherwise. Note that this procedure is very similar to M-estimators (such as the Huber estimator) for robust class means \cite{Huber2009}. The only difference is that the Huber estimator computes a weighted linear combination of the sample points, whereas we use a `harder' form of weights, i.e. we discard the outliers entirely.

\subsection{Patch-Based Denoising}
\label{subsec:denoising}
The processed cluster centers $\{\tilde{p}_j\}_{j=1}^{K_c}$ as obtained in the previous step are expected to be devoid of outliers. The averaging also induces a basic form of filtering to remove the noise. However, some residual noise still remains. Hence, we pass these cluster centers ($\{\bar{p}_j\}_{j=1}^{K_c}$) through a denoising algorithm described next. We use a patch-based PCA denoising method to reduce the noise in the projections. This algorithm is adapted from a similar algorithm for 2-D images, as described in \cite{MuresanAdaptiveDenoising}. In this algorithm, we extract small-sized patches from each cluster center. For each such patch, we find some $L$ patches nearest to it in terms of the $\ell_2$ distance. After performing PCA on this set of patches, we project each patch along the principal direction to produce eigen-coefficients. To denoise the patch, we manipulate these coefficients using Wiener-like updates of the form
\begin{equation}
\beta_{il} = \alpha_{il} \Bigg(\frac{\sigma_l^2}{\sigma_l^2 + \sigma^2/\bar{K}}\Bigg)
\end{equation}
where $\beta_{il}$ is an estimate of the $l^{th}$ denoised coefficient for patch $i$ (part of cluster center $\bar{p}_j$), $\alpha_{il}$ is the corresponding noisy coefficient, $\sigma^2$ is the noise variance in the original projections which is assumed to be known, $\bar{K}$ is the average number of projection vectors assigned to a cluster, and $\sigma_l^2$ is the mean square value of the $l^{th}$ coefficient estimated as follows:
\begin{equation}
\sigma_l^2 = \max\Bigg(0, \frac{1}{L}\sum_{i=1}^L \alpha_{il}^2 - \sigma^2/\bar{K}\Bigg).
\end{equation} 
This patch-based PCA denoising algorithm is better than the PCA denoising algorithm used in \cite{Singer2013Two-DimensionalDirections}. This is because, in \cite{Singer2013Two-DimensionalDirections}, entire projections are compared instead of just patches. The advantage of our patch-based approach is that it is easier to find a number of patches which are structurally similar to a given reference patch, but that is not true for entire projections. The second advantage is that we now have the freedom to find similar patches from within a projection vector, but from other projection vectors as well. Moreover, we performed the denoising in sliding window fashion with a pixel stride of 1, resulting in several potential denoised values per pixel. The final denoised value was selected using averaging. Hereafter, we use the symbol $\tilde{q}_i$ to refer to the denoised version of the cluster center $\tilde{p}_i$ after the outlier removal step.

\subsection{Initialization of the orientations and shifts using Helgason Ludwig Consistency Conditions (HLCC)}
\label{subsec:hlcc}
Determining the object structure from projections with unknown viewing parameters is a highly non-convex optimization problem. Starting from a random initialization of orientations, the algorithm may mostly converge to a local optimum and fail to give us the right structure. This is why a good initialization of the orientations and the shifts is necessary. Initially, we attempted to correct the shifts by shifting all the projections such that their center of mass is at the origin as stated in \cite{Basu2000UniquenessAngles}. In practice, however, in spite of correcting a few shifts, it still resulted in unsatisfactory reconstruction. So we modified our approach and harnessed the information available in the image moments and projection moments to estimate the shifts and the orientations simultaneously. The HLCC \cite{Natterer2001TheTomography} give us a relationship between the geometric moments of the underlying image $z(x, y)$ and those of its unshifted projections at any angle. We use this to derive a good initial estimate of the unknown orientations and shifts. The basis behind the algorithm is as follows.
\noindent
The moments of order $p,q$ of the image $z(x,y)$ are given by
\begin{equation}
v_{p, q} = \int_{-\infty}^{\infty}\int_{-\infty}^{\infty} z(x, y) x^p y^q dx dy.
\end{equation}
The $n^{\textrm{th}}$ order moment of the projection $g(\rho, \theta) \triangleq \int_{-\infty}^{\infty} z(x,y) \delta(\rho - x \cos \theta - y \sin \theta) dx dy$ is given by 
\begin{equation} \label{proj_moment}
m_{\theta}^{(n)} = \int_{-\infty}^{\infty} g(\rho, \theta) \rho^n d\rho.
\end{equation}
\noindent
If the projection is shifted by $s_i$ to give a projection $g(\rho, \theta, s_i)$, its $n^{th}$ order moment after reverse shifting by an amount $s_k$ can be written as
\begin{equation}
m_{\theta, s_k}^{(n)} = \int_{-\infty}^{\infty} \mathcal{S} \{g(\rho, \theta, s_i), s_k\} \rho^n d\rho
\end{equation}
\noindent
where $\mathcal{S} \{., s_k\}$ denotes the reverse shift operation. The above evaluates to the same quantity as (\ref{proj_moment}) if $s_k = s_i$. That is,
\begin{equation}
m_{\theta, s_i}^{(n)} = \int_{-\infty}^{\infty} \mathcal{S} \{g(\rho, \theta, s_i), s_k\} \rho^n d\rho = m_{\theta}^{(n)}.
\end{equation}
\noindent
The HLCC give a relationship between $m_{\theta, s_i}^{(n)}$ and $v_{p, q}$, which is defined as
\begin{equation}
m_{\theta, s_i}^{(n)} = \sum_{j=0}^{n} \binom{n}{j} (cos\theta)^{n-j}(sin \theta)^j v_{n-j, j}.
\label{eq:hlcc}
\end{equation}
\noindent
Thus for each order $n$, we can write the constraints in matrix form, $\textbf{m}^{(n)} = \textbf{A}^{(n)}\textbf{v}^{(n)}$. Here, for a total of $K_c$ projections and for the $n^{th}$ order equation, $\textbf{A}^{(n)}$ is the $K_c\times(n+1)$ matrix defined by $\textbf{$A_{ij}^{(n)}$} = \binom{n}{j} (cos\theta_i)^{n-j}(sin \theta_i)^j$, and $\textbf{$v^{(n)}$} \triangleq \{\textbf{$v_{p, q}$} | (p + q) = n, p,q \in \mathbb{Z}_{\geq 0}\}$. 
\noindent
Since, in practice, the projections are noisy and the shifts unknown, Eqn. \ref{eq:hlcc} will not be satisfied exactly. Instead we define an energy function as follows
\begin{equation}
E(\{\theta_i\}, \textbf{v}, \{s_i\}) = \sum_{n=0}^k \sum_{i=1}^{K_c} \Bigg ( m_{\theta_i, s_i}^{(n)} - \sum_{j=0}^n A_{i, j}^{(n)} v_{n-j, j} \Bigg )^2.
\end{equation}
Note that in this equation, the moments $m_{\theta_i, s_i}^{(n)}$ correspond to those of the $i^{th}$ cluster center $\tilde{q}_i$ (post-denoising) where $1 \leq i \leq K_c$.
By minimizing this energy function, we derive an initial estimate of the angles and the shifts through an iterative coordinate descent strategy as in \cite{Malhotra2016TomographicRelationships}. A small number of multi-starts (around 10), each with a different random initialization of the pose parameters, helped further combat the non-convexity of the objective function $E(\{\theta_i\}, \textbf{v}, \{s_i\})$, and in fact yielded much superior results. In case of multiple starts, the solution which yielded the least value of the objective function, was selected.

\subsection{Optimization strategy to obtain the structure of the object}
\label{subsec:opt}
After clustering and obtaining an initial estimate of the angles and shifts, we initially decided to use a sparsity based optimization technique due to the promising results delivered by compressed sensing \cite{Wang2010}. The following optimization problem was selected:
\begin{equation}
\mathcal{L} (\{\theta_i\}, \beta, \{s_i\}) = \sum_{i=1}^{K_c} \|\tilde{q}_{i, s_i} - \mathcal{R}_{\theta_i} (U\beta)\|^2_2 + \lambda_1\|\beta\|_1
\end{equation}
\noindent 
where $\{\theta_i \}_{i=1}^{K_c},\{s_i \}_{i=1}^{K_c}$ are the $K_c$ unknown angles and shifts for the cluster centers $\{\tilde{q}\}_{i=1}^{K_c}$, $\tilde{q}_{i, s_i}$ denotes $\tilde{q}_i$ shifted by $s_i$, the matrix $\mathcal{R}_{\theta_i}$ represents the set of line integrals at different shifts along the direction $\theta_i$ (for computing the Radon transform), $U$ denotes the inverse discrete cosine transform (DCT) operator or any other sparsifying operator, and $\beta$ is the vector of DCT or other transform coefficients of the image to be reconstructed. That is, the image is represented as $z = U \beta$, where $\beta$ is a sparse or compressible vector of transform coefficients. The function was minimized using an alternating method over the unknown angles $\{\theta_i \}_{i=1}^{K_c}$, the unknown shifts $\{s_i \}_{i=1}^{K_c}$, and image DCT coefficients $\beta$. \\
\noindent
Solving this problem without the initial estimates is extremely ill-posed because of the large number of degrees of freedom. Therefore, we hypothesized that the initial estimates provided by the moment-based estimation would serve as a good initial estimate for this problem and we would converge onto the accurate structure of the object. However, upon solving this optimization problem, the obtained results exhibited severe artifacts, as we show in Section \ref{sec:results}. We also noticed convergence problems. If the accurate angles and shifts were provided, the sparsity-based technique, however, provided excellent reconstruction and no convergence problems arose. Hence we concluded that this reconstruction method is indeed very sensitive to errors in the angles, and even the initial estimate from the moment based approach was not accurate enough for this optimization to converge. 
\\ \noindent
\textbf{Back to FBP:} In light of the above observations, we decided to consider an alternative optimization problem which we observed was significantly more robust to errors in the initial estimate. We hence considered the following optimization problem, and decided to minimize the following function using a stochastic gradient descent approach: 
\begin{equation}
\mathcal{M} (\{\theta_i\}, z, \{s_i\}) = \sum_{i=1}^{K_c} \|\tilde{q}_{i, s_i} - \mathcal{R}_{\theta_i} (z)\|^2_2.
\end{equation}
Given an initial estimate of the angles and the shifts, the gradient with respect to the image $z$, is given by
\begin{equation}
\nabla_z \mathcal{M} (\{\theta_i\}, z, \{s_i\}) = \sum_{i=1}^{K_c} -2\mathcal{R}_{\theta_i}^T (\tilde{q}_{i, s_i} - \mathcal{R}_{\theta_i}(z)),
\end{equation}
where $\mathcal{R}_{\theta_i}^T$ is the adjoint operator  for $\mathcal{R}_{\theta_i}$. Using the initial estimate of the orientations and the shifts provided by the moments based approach, the problem was solved in an alternating way. We first estimated the structure using the gradient calculated above, which effectively makes use of FBP-based reconstruction. Given an estimate of the structure,  the orientation and the shift of each projection by coordinate descent with a single-dimensional brute-force search. The complete procedure is summarized below in Algorithm 1.
\begin{algorithm}
\caption{Complete algorithm}
\begin{algorithmic}
\State ${C_i} \gets \text{ Cluster assignment of the } i^{th} \text{ projection}$
\State $\xi_j \gets \text{ Centroid of the } j^{th} \text{ cluster}$
\State $I_i \gets \text{Indicator of } i^{th} \text{ projection being filtered or not}$\\

\State $ \{C_i\}, \{\xi_j\} = Lq\_KMeans(\{y_i\}), 0 < q <= 1$
\State $\{I_i\} = Outlier\_Detection(\{y_i\}, \{C_i\}, \{\xi_j\})$
\State $\{\tilde{p}_i\} = Robust\_Average\_Projections(\{y_i\}, \{C_i\}, \{\pi_j\}, \{I_i\})$
\State $\{\tilde{q}_i\} = Denoise\_Projections(\{\tilde{p}_i\})$
\State $\{\theta_i\}, \{s_i\} = MomentsBasedSolver(\{\tilde{q}_i\}, k)$ \\

\State $\alpha \gets LearningRate$
\State $E \gets \infty$
\State $\Delta E \gets \infty$ \\

\While{$\Delta E > \tau$}
\State $\text{Choose a random subset of projections of size } K' < K_c$
\State $\{\tilde{q}_l\}_{l=1}^{K'} $\\

\State Update the image using the $K'$ projections
\State $ z \gets z - \alpha(z - FBP(\{\tilde{q}_l\}, \{\theta_l\}, \{s_l\}))$ \\

\State Refine all the $K_c$ angle and shift estimates
\For{$i=1:K_c$}
\State $\{\theta_i\}, \{s_i\} = Best\_Theta\_And\_Shift(z, \{\tilde{q}_i\})$
\EndFor
\State \textbf{end for} \\

\State $\alpha \gets \alpha - AnnealRate$
\State $\Delta E = E - \sum_{i=1}^{K_c} \|\tilde{q}_{i,s_i} - \mathcal{R}_{\theta_i} (z)\|^2_2$
\State $E = \sum_{i=1}^{K_c} \|\tilde{q}_{i, s_i} - \mathcal{R}_{\theta_i} (z)\|^2_2$
\EndWhile
\State \textbf{end while}
\end{algorithmic}
\end{algorithm}

\section{Results}
\label{sec:results}
\begin{figure}[H]
\begin{subfigure}{0.23\textwidth}
\centering
\includegraphics[width=0.7\linewidth]{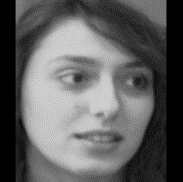}
\caption{Original image}
\end{subfigure}
\begin{subfigure}{0.23\textwidth}
\centering
\includegraphics[width=0.7\linewidth]{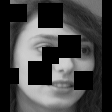} \caption{Class 2 outlier}
\end{subfigure}
\begin{subfigure}{0.23\textwidth}
\centering
\includegraphics[width=0.7\linewidth]{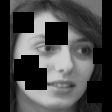}
\caption{Class 2 outlier}
\end{subfigure}
\begin{subfigure}{0.23\textwidth}
\centering
\includegraphics[width=0.7\linewidth]{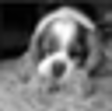} 
\caption{Class 1 outlier}
\end{subfigure}
\caption{Original image, and images which generated Outliers of class 1 and class 2}
\label{fig:outliers}
\end{figure}

\begin{figure}[H]
\includegraphics[width=0.5\textwidth, height=4.8cm]{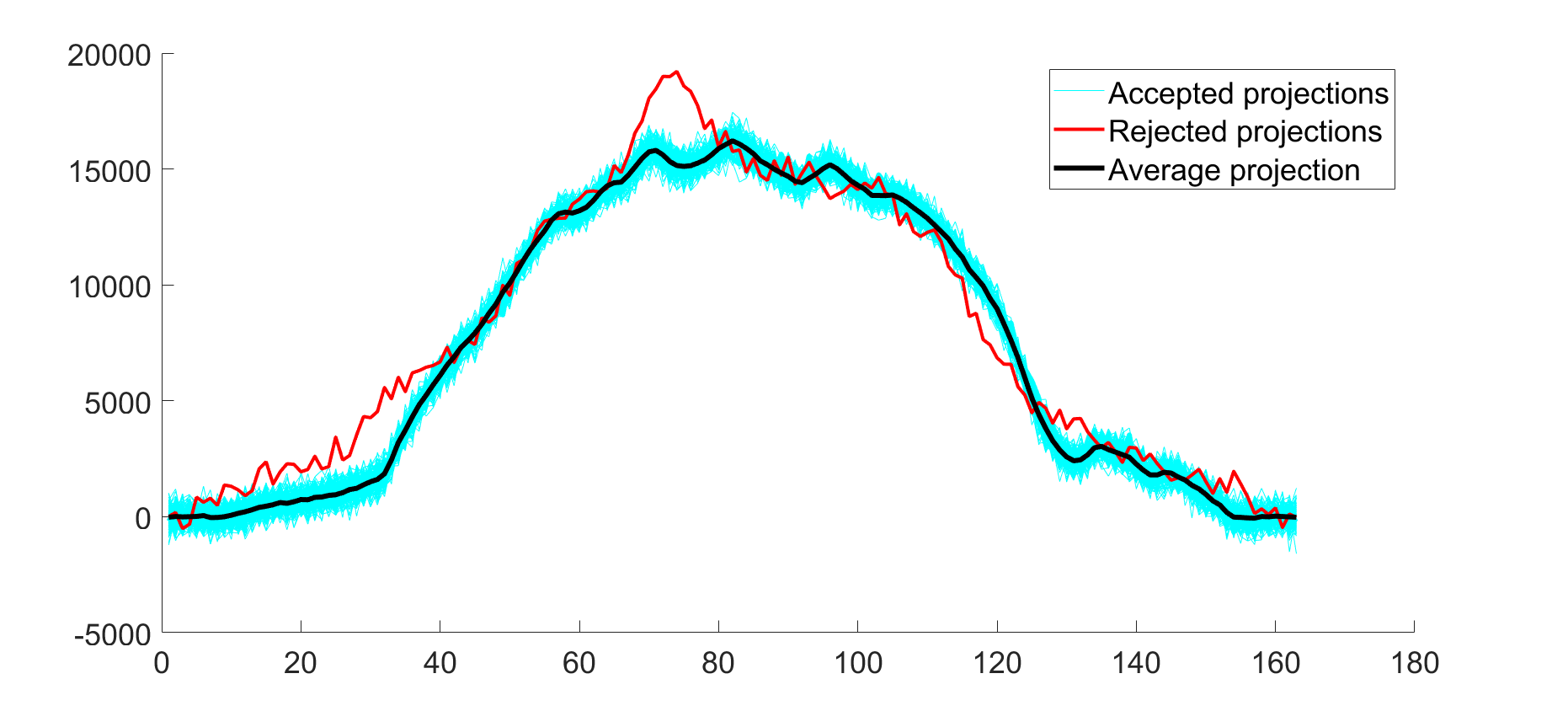} 
\caption{The red projection belonging to the class 1 outlier is filtered out in this step.}
\label{fig:projections}
\end{figure}

\noindent
In this section, we present a complete set of results on the algorithm described in the previous section. Using our algorithm, we demonstrate how to successfully tackle all the problems mentioned earlier and achieve good quality reconstructions of the original object. The images used for our experiments were taken from the Yale and ORL face databases and the image sizes used were $192 \times 192$ and $112 \times 112$ respectively. A total of $Q = 2 \times 10^4$ projections per image were simulated using angles from $\textrm{Uniform}(0,\pi)$\footnote{Though we considered the Uniform distribution, our algorithm does not rely on this assumption, or knowledge of the distribution of the orientations.}. A fraction $f_1$ of these projections were outliers of class 1, i.e. they were projections of non-face images taken from the CIFAR-10 dataset \cite{Krizhevsky2009LearningImages}.  For another fraction $f_2$ of projections, we deliberately generated them from a copy of the same image, but with a small number ($f_3\%$) of pixel values (at randomly selected locations) set to 0. We term the corresponding projections `Outliers of Class 2'. These simulate projections of biological specimens corrupted by overlapping ice particles or minor structural changes. Some sample illustrative images are presented in Fig. \ref{fig:outliers}. All projections were subjected to additive i.i.d. noise from $\mathcal{N}(0,\sigma^2)$, where we assume $\sigma$ to be known in advance, even though there are techniques to estimate it directly from the noisy projections. The noisy projections were clustered into $K_c = 180$ angles. The outlier projections were then identified and removed using the procedure described earlier in Section \ref{subsec:class1}. 
\noindent
We observed that in many cases, the outlier projections lying farther away from the cluster centroid were filtered out, and only the true projections of the object were preserved for the consequent steps. Fig. \ref{fig:projections} shows an example of outlier and noisy inlier projections as well as the average projection vector after removing outlier projections. This example is for a face image from the ORL database, with outliers from the CIFAR dataset. The relevant parameters are $f_1 = 10\%, f_2 = 10\%, f_3 = 10\%$ and $\sigma = 0.1 \times$ average value of noiseless projections.

\begin{figure}[H]
\begin{subfigure}{0.5\textwidth}
\centering
\includegraphics[width=1.0\linewidth, height=4.8cm]{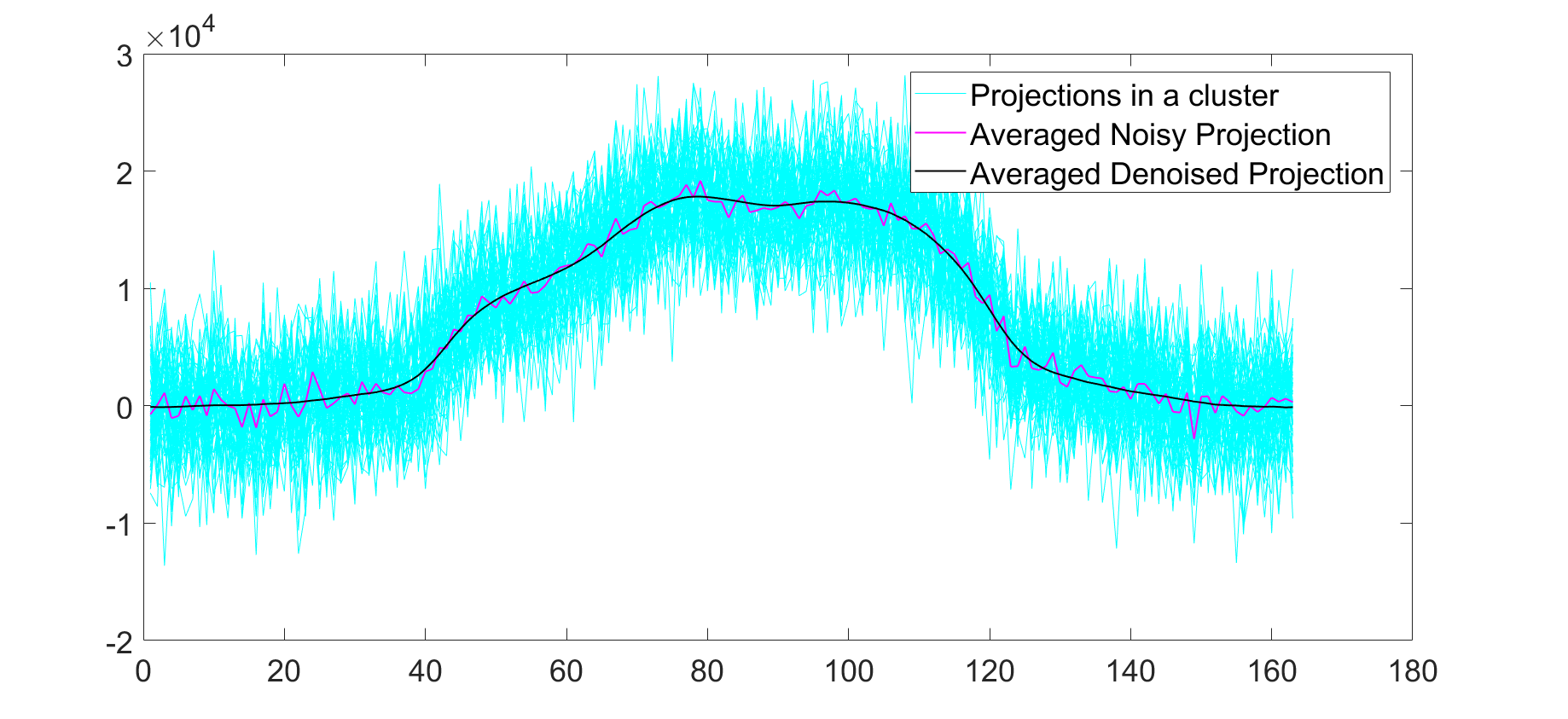} 
\caption{Denoising Example 1, 50\% noise}
\end{subfigure}
\begin{subfigure}{0.5\textwidth}
\centering
\includegraphics[width=1.0\linewidth, height=4.8cm]{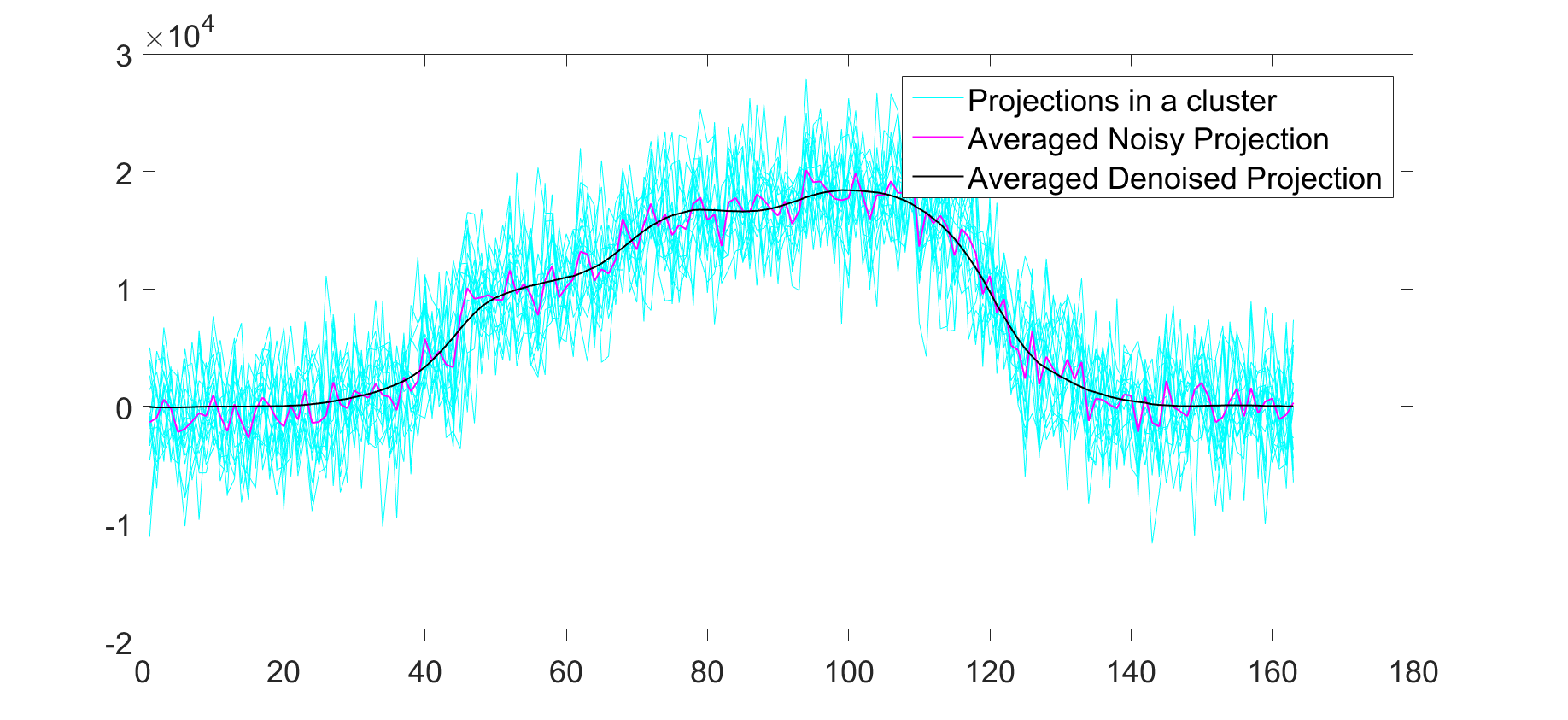}
\caption{Denoising Example 2, 50\% noise}
%
\end{subfigure}
\caption{Patch-based denoising for two different clusters}
\label{fig:denoising_projection}
\end{figure}

\begin{figure}[H]
\includegraphics[width=1.0\linewidth, height=5cm]{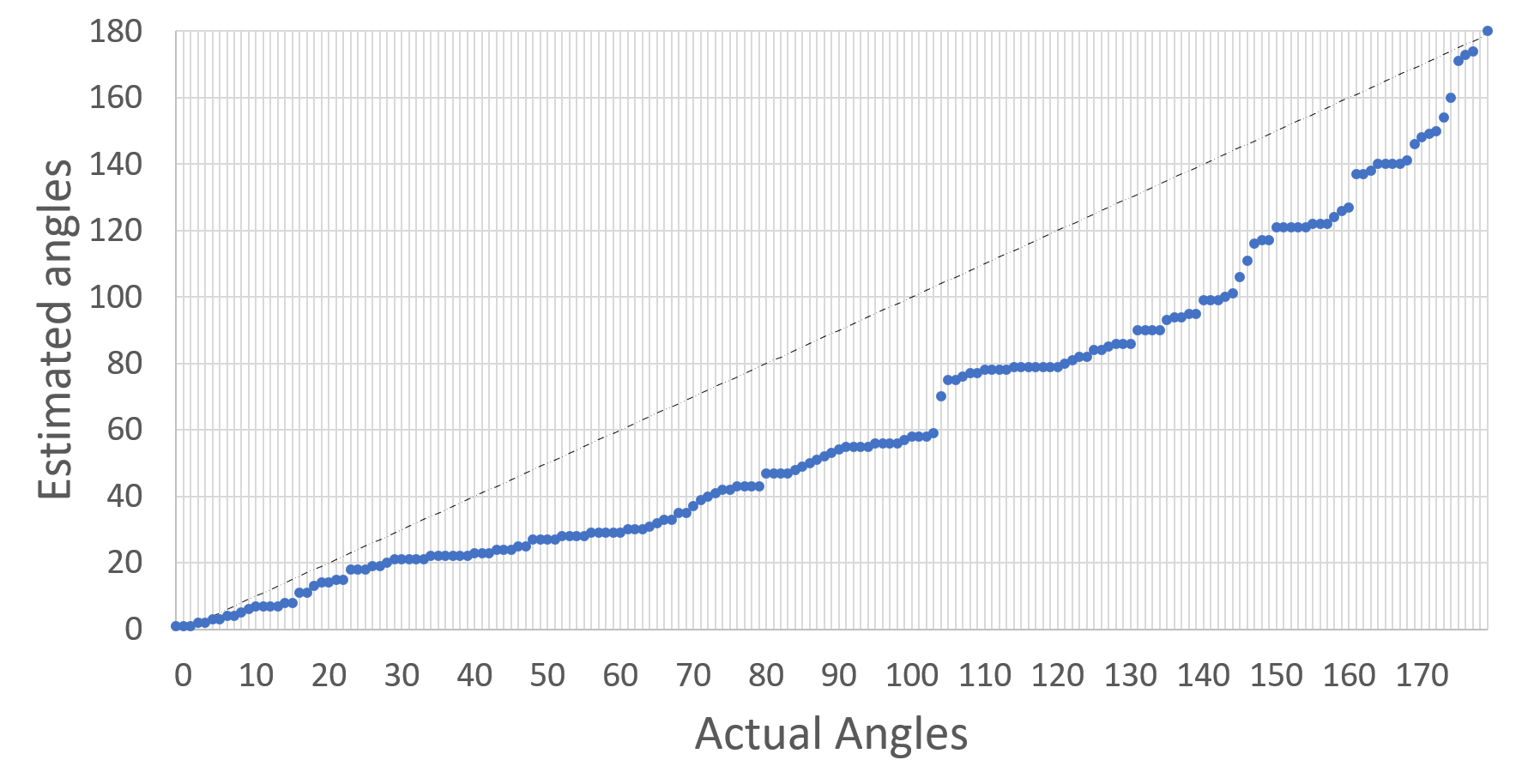} 
\caption{Initial estimate of orientations: $\sigma = 10\%$ Noise, $f_1 = 10\%$ outliers of class 1, $f_2 = 10\%$ outliers of class 2}
\label{fig:moment_orientations}
\end{figure}

\begin{figure}[H]
\begin{subfigure}{0.23\textwidth}
\centering
\includegraphics[width=0.7\linewidth]{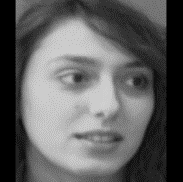} 
\caption{Original Image}
\end{subfigure}
\begin{subfigure}{0.23\textwidth}
\centering
\includegraphics[width=0.7\linewidth]{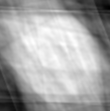}
\caption{Reconstructed Image}
\end{subfigure}
\caption{An example of the failure of the moments based approach for image reconstruction. Parameters: 10\% Noise, 10\% outliers of class 1, 10\% outliers of class 2}
\label{fig:badrecon}
\end{figure}

\begin{figure}[H]
\begin{subfigure}{0.23\textwidth}
\centering
\includegraphics[width=0.7\linewidth]{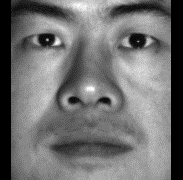} 
\caption{Original Image}
\end{subfigure}
\begin{subfigure}{0.23\textwidth}
\centering
\includegraphics[width=0.7\linewidth]{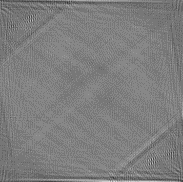}
\caption{Reconstructed Image}
\end{subfigure}
\caption{An example of the failure of sparsity-based optimization, with angle errors}
\label{fig:csrecon} 
\end{figure}

\vspace{0.1cm}
\noindent
After outlier removal, the remaining projections in each cluster were averaged and passed on to the patch-based denoising step. An example of the projections after the denoising step are shown in Fig. \ref{fig:denoising_projection}. This is for a noise level of $\sigma = 0.5 \times$ average value of noiseless projections.
\vspace{0.07cm}
\noindent
The denoised projections were passed to the moments-based solver for an initial estimate of the orientations. However the problem of tomography under unknown angles is inherently ambiguous and the solutions can be obtained only up to a global rotation. Hence the estimated orientations will in most cases be shifted from the original values by a \emph{single global} offset $\delta$. To quantify the accuracy of the estimates, the orientation estimates obtained using the HLCC-based method from Section \ref{subsec:hlcc} were corrected for this global rotational ambiguity and then displayed in Fig. \ref{fig:moment_orientations}. In Fig. \ref{fig:moment_orientations}, we show a scatter plot of the $K_c$ projection angles corresponding to `ground-truth cluster centers' (i.e. cluster centers obtained from noiseless projection vectors without outliers) and the corresponding values of the estimates of those angles after correcting for the offset $\delta$. The ground truth angles turn out to have values from 0 to $180^{\circ}$, because of the uniform distribution of the orientations of the projection vectors.
\noindent
A perfect set of estimates would have produced a plot aligned with the $45^{\circ}$ line. However, as seen in Fig. \ref{fig:moment_orientations}, the initial estimate provided by the moments based approach deviated significantly from the actual orientations (even after correction for the global offset). Reconstruction of the object considering these orientations as the true values yielded poor results as seen in Fig. \ref{fig:badrecon}.
\noindent
This showed that there was a clear necessity to refine the results given by the moments based approach. We therefore explored two ways to potentially improve the estimate obtained from the moment-based initialization method - the sparsity-based optimization framework, and the gradient descent based approach - both described in Section \ref{subsec:opt}. The former was implemented using the $\ell_1-\ell_s$ package \cite{Kim2007AnSquares}. As explained in Section \ref{subsec:opt}, the sparsity-based reconstruction quite surprisingly failed to yield good results, an example of which is seen in Fig. \ref{fig:csrecon}. This was mainly due to errors in orientation estimates. 
\subsection{Reconstruction using the gradient descent based approach}
\begin{figure}[H]
\begin{subfigure}{0.23\textwidth}
\centering
\includegraphics[width=0.7\linewidth]{images/person1} 
\caption{Original Image}
\end{subfigure}
\begin{subfigure}{0.23\textwidth}
\centering
\includegraphics[width=0.7\linewidth]{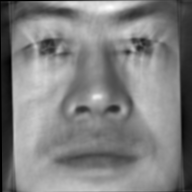}
\caption{Reconstructed Image}
\end{subfigure}
\caption{Results of Algorithm 1 with 0\% Noise and no outliers of class 1 or class 2, RMSE - 4.58\%}
\label{fig:no_noise}
\end{figure}

After the failure of the sparsity-based optimization framework, we altered our approach and adopted a gradient-based optimization framework using the FBP, and the results turned out to be very promising. Since the problem is ambiguous up to a global rotation, the reconstructed image might be a rotated version of the original image. Therefore the reconstruction was registered to the original test image, to obtain a quantitative error metric. The error metric used was the Relative Mean Squared Error between the registered reconstruction and
the test image. This is defined as follows: $\textrm{RMSE}(z,\hat{z}) = \dfrac{\|z-\hat{z}\|_2}{\|z\|_2}$ where $\hat{z}$ is the reconstructed estimate for $z$. Reconstruction without any noise and outliers produced nearly accurate results, an example of which can be seen in Fig. \ref{fig:no_noise}.

\begin{figure}[H]
\begin{subfigure}{0.23\textwidth}
\centering
\includegraphics[width=0.7\linewidth]{images/person2} 
\caption{Original Image}
\end{subfigure}
\begin{subfigure}{0.23\textwidth}
\centering
\includegraphics[width=0.7\linewidth]{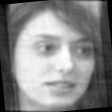}
\caption{Reconstructed Image}
\end{subfigure}
 \caption{10\% Noise, 10\% outliers of class 1, 10\% outliers of class 2, RMSE - 8.95\%}
 \label{fig:refined_recon}
\end{figure}

\begin{figure}[H]
\includegraphics[width=1.0\linewidth, height=5cm]{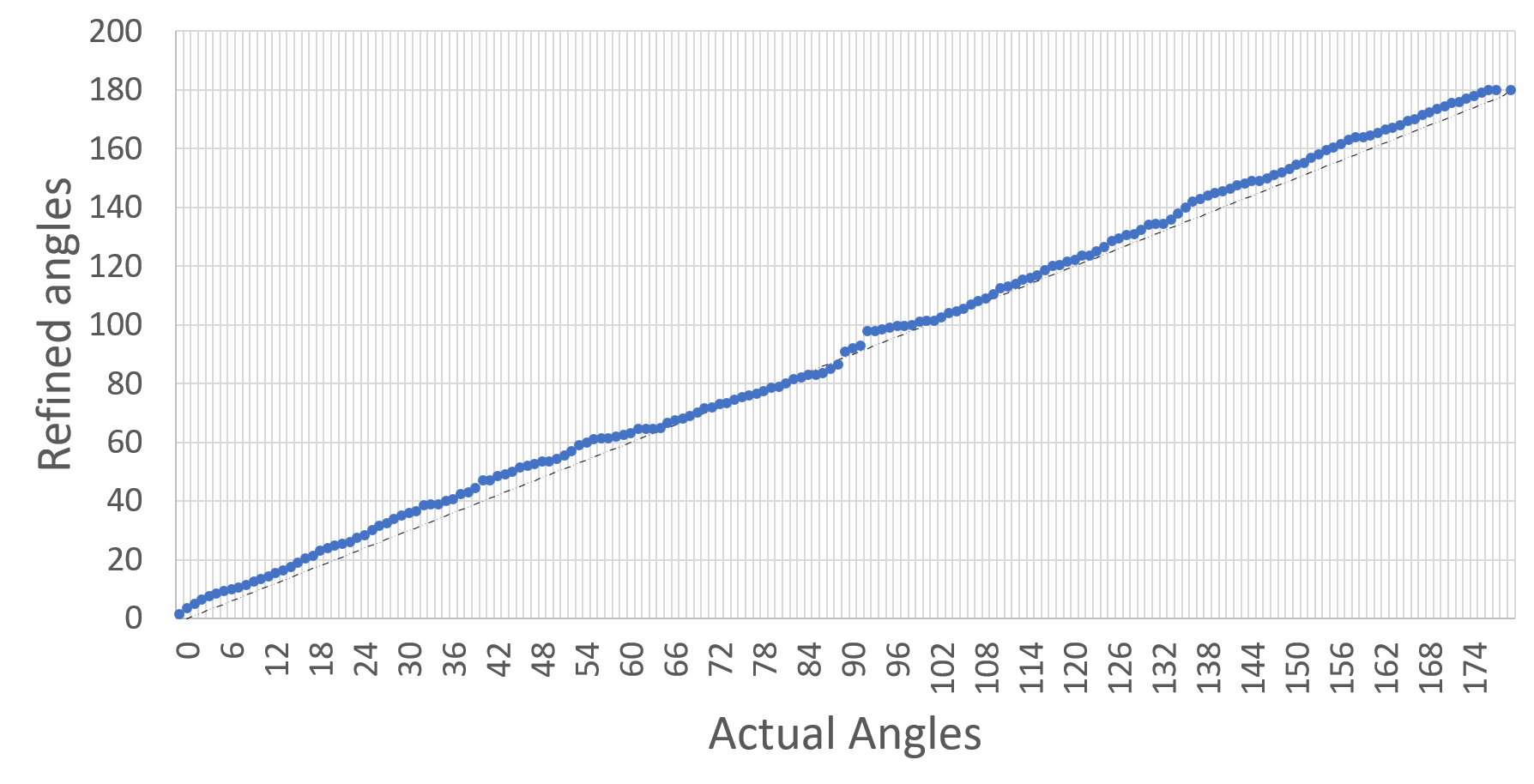} 
\caption{Refined estimate of orientations: 10\% Noise, 10\% outliers of class 1, 10\% outliers of class 2}
\label{fig:refined_angles}
\end{figure}

\noindent
However even in cases of noisy projections with 10\% noise, and a significant percentage of outliers of class 1 and class 2 (10\% each), we observed that our algorithm was able to successfully refine the initial estimate given by the moments based approach and obtain an accurate reconstruction. On applying our algorithm to the moment-based orientation estimates that yielded the results shown in Fig. \ref{fig:moment_orientations} and \ref{fig:badrecon}, we obtained significantly refined angle estimates, as shown in Fig. \ref{fig:refined_angles}. A sample reconstruction is shown in Fig. \ref{fig:refined_recon}.

\begin{figure}[H]
\begin{subfigure}{0.23\textwidth}
\centering
\includegraphics[width=0.7\linewidth]{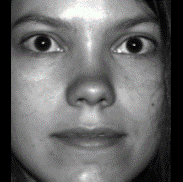} 
\caption{Original Image}
\end{subfigure}
\begin{subfigure}{0.23\textwidth}
\centering
\includegraphics[width=0.7\linewidth]{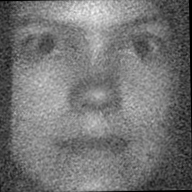}
\caption{Reconstructed Image}
\end{subfigure}
\caption{50\% Noise, 0\% outliers of class 1, 0\% outliers of class 2, RMSE - 11.99\%}
\label{fig:recon_high_noise1}
\end{figure}

\begin{figure}[H]
\begin{subfigure}{0.23\textwidth}
\centering
\includegraphics[width=0.7\linewidth]{images/person2} 
\caption{Original Image}
\end{subfigure}
\begin{subfigure}{0.23\textwidth}
\centering
\includegraphics[width=0.7\linewidth]{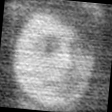}
\caption{Reconstructed Image}
\end{subfigure}
\caption{50\% Noise, 5\% outliers of class 1, 5\% outliers of class 2, RMSE - 18.39\%}
\label{fig:recon_high_noise2} 
\end{figure}

\noindent
To check the robustness of the algorithm, we also experimented with reconstructions under higher noise levels and a higher percentage of outliers. The reconstructions in Fig. \ref{fig:recon_high_noise1} and \ref{fig:recon_high_noise2} were obtained for extremely high noise levels (50\% of the average value of the projections).

\begin{figure}[H]
\includegraphics[width=1.0\linewidth, height=5cm]{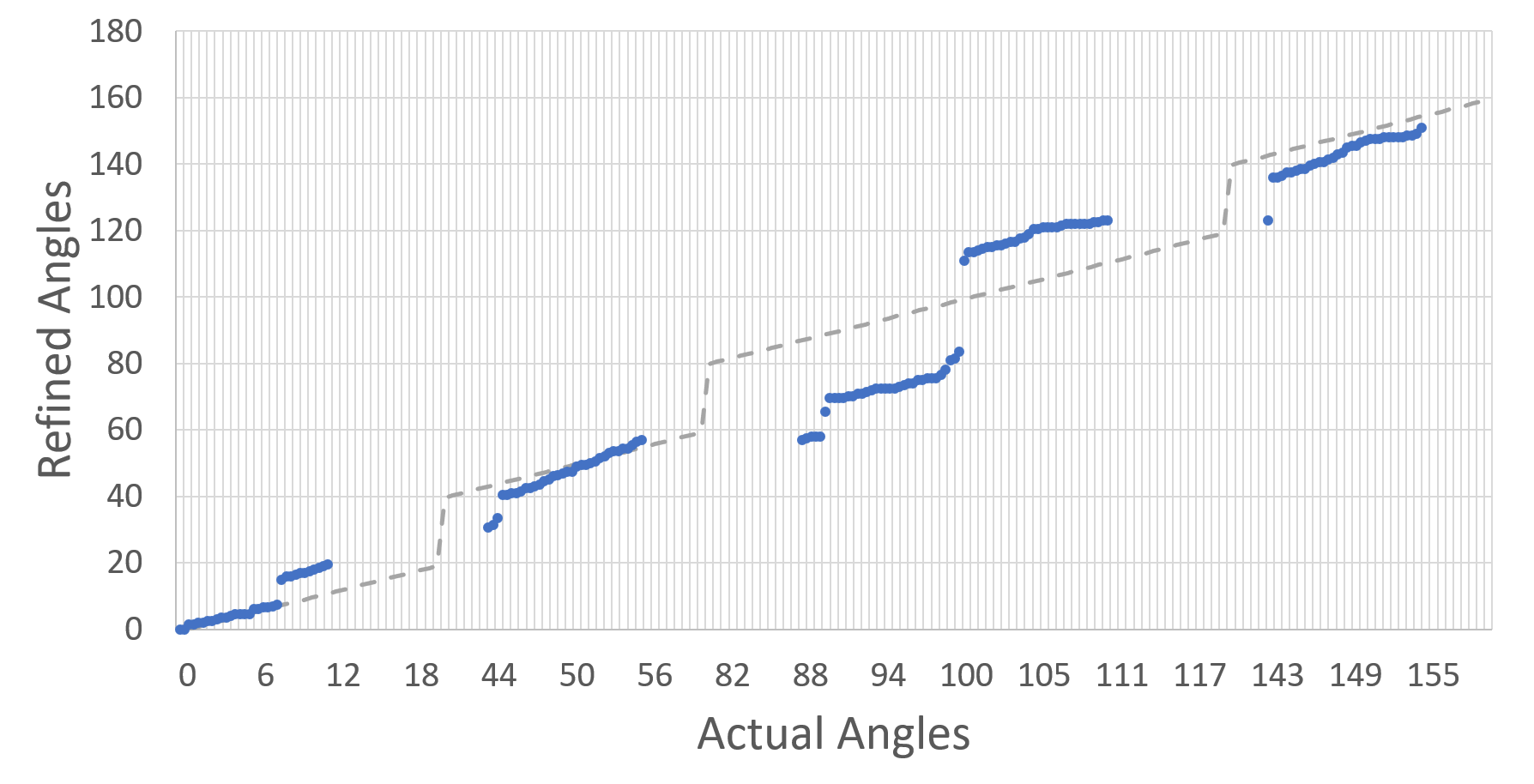}
\caption{Refined estimate of orientations in case of non-uniform distribution of angles: 20\% Noise, 10\% outliers of class 1, 0\% outliers of class 2}
\label{fig:refined_non_uniform_angles}
\end{figure}

\begin{figure}[H]
\begin{subfigure}{0.23\textwidth}
\centering
\includegraphics[width=0.7\linewidth]{images/person2} 
\caption{Original Image}
\end{subfigure}
\begin{subfigure}{0.23\textwidth}
\centering
\includegraphics[width=0.7\linewidth]{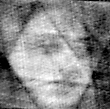}
\caption{Reconstructed Image}
\end{subfigure}
\caption{20\% Noise, 10\% outliers of class 1, 0\% outliers of class 2, Non-Uniform distribution of angles, RMSE - 17.69\%}
\label{fig:recon_non_uniform} 
\end{figure}

\noindent
\textbf{Experiments with non-uniform orientation distributions:} Unlike previous algorithms, our algorithm doesn't assume that the orientations of the projections are uniformly distributed, nor does it require knowledge of the underlying distribution. To test this, we conducted the following experiment. Instead of angles being taken from the $\textrm{Uniform}(0,\pi)$ distribution, we considered the following distribution for the projection orientations: 
$\textrm{Uniform}(0,\pi/9) \cup \textrm{Uniform}(2\pi/9,\pi/3) \hspace{0.1cm}\cup
       \textrm{Uniform}(4\pi/9,2\pi/3) \cup \textrm{Uniform}(7\pi/9,8\pi/9). 
$\\
Further to re-emphasize the robustness of our algorithm, the projections were subjected to high amounts of noise (20\% of the average value of the projections) and 10\% outliers of class 1. The distribution of the original angles along with our estimates of the angles after the moment based solver and refinement are shown in Fig. \ref{fig:refined_non_uniform_angles}. The final reconstruction is shown in Fig. \ref{fig:recon_non_uniform}. Note that in generating this result, the algorithm did not exploit any knowledge whatsoever of the distribution of the orientation. 

\begin{figure}[H]
\begin{subfigure}{0.45\textwidth}
\centering
\includegraphics[width=1.0\linewidth, height=4.5cm]{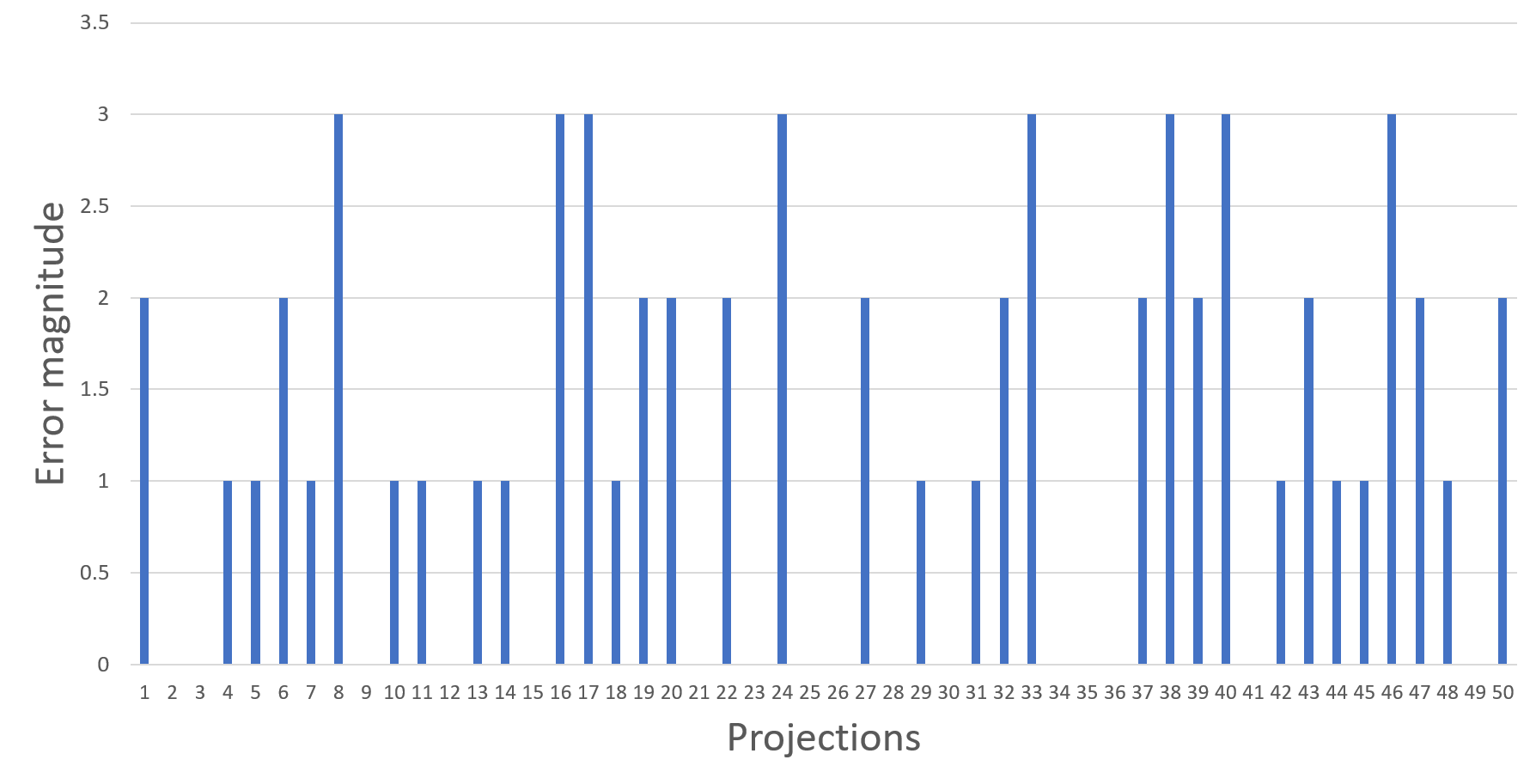} 
\caption{Shifts estimated by the moment based approach}
\end{subfigure}
\begin{subfigure}{0.45\textwidth}
\centering
\includegraphics[width=1.0\linewidth, height=4.5cm]{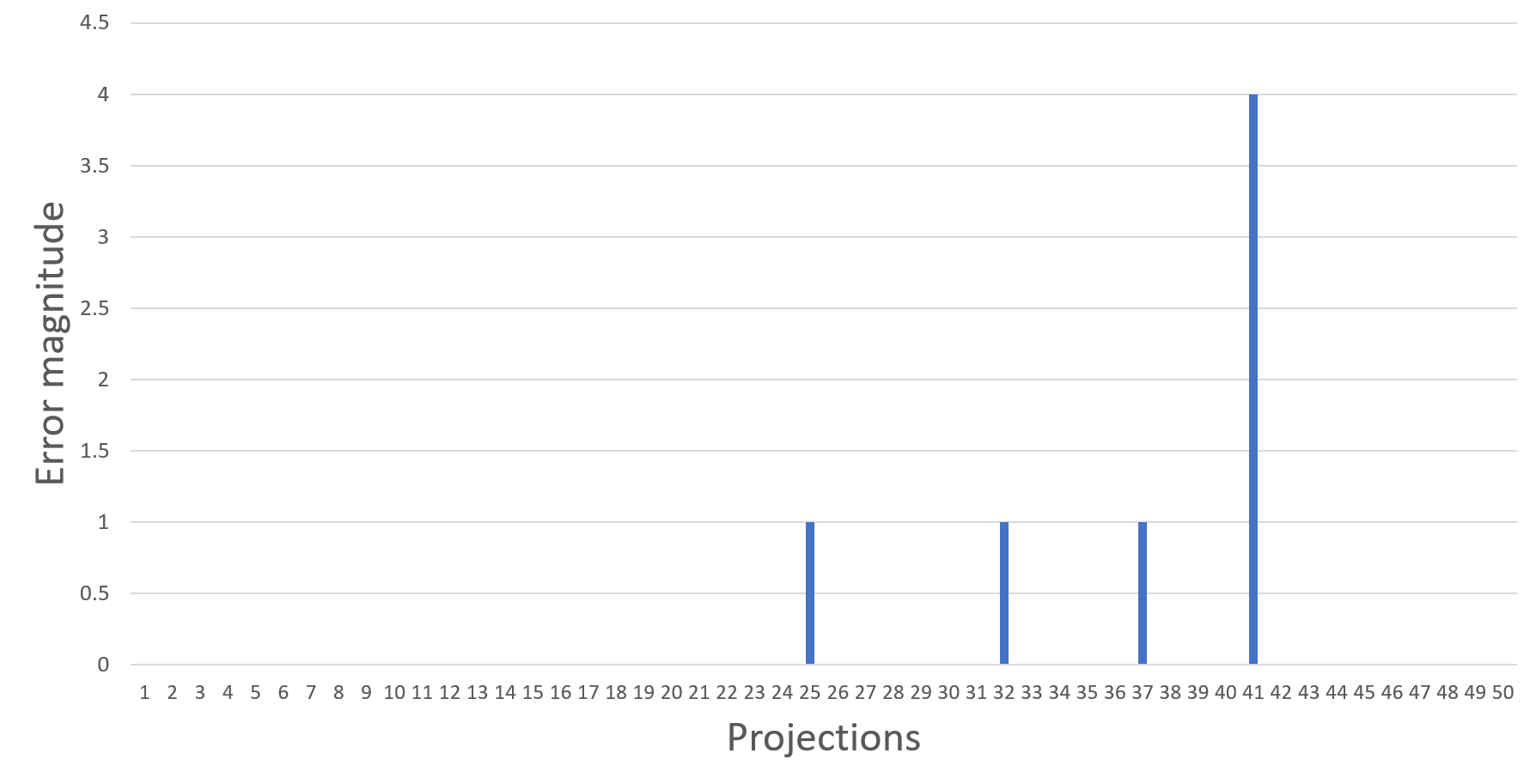}
\caption{Refined estimates of the shifts}
\end{subfigure}
\caption{Estimation of shifts}
\label{fig:shift_estimate}
\end{figure}

\begin{figure}[H]
 \begin{subfigure}{0.23\textwidth}
\centering
\includegraphics[width=0.7\linewidth]{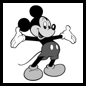} 
\caption{Original Image}
\end{subfigure}
\begin{subfigure}{0.23\textwidth}
\centering
\includegraphics[width=0.7\linewidth]{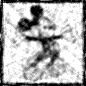}
\caption{Reconstructed Image}
\end{subfigure}
\caption{10\% Noise and unknown shifts of maximum amplitude $\pm 2$, RMSE-5.65\%}
\label{fig:shift_reconstruction}
\end{figure}

\noindent
\textbf{Experiments with unknown shifts:} In cases where projections have random unknown shifts, our algorithm was able to accurately estimate the shift in each projection and correct it to produce an accurate reconstruction. The following has been tested on a $86 \times 86$ image, with 50 projections, 10\% Noise and random unknown shifts up to $\pm 2$. Note that the limit of $\pm2$ on the shifts is not overly restrictive, as indeed the range of possible shifts is very small in these applications. The plot in Fig. \ref{fig:shift_estimate} shows the absolute error between the actual shifts in the projection and the shifts we estimated (after correction for an inevitable global shift ambiguity). A sample reconstruction is shown in Fig. \ref{fig:shift_reconstruction}

\section{Discussion and Conclusion}
\label{sec:typestyle}
From the results presented here, we conclude that the algorithm described in this paper is capable of estimating the original underlying structure to a high degree of accuracy, without any prior knowledge about the angles of projections and any prior structural information such as templates. Further, we have also seen that our method is robust to the presence of outliers, projections with extremely high amounts of noise and projections with random unknown shifts. Our method does not make any assumption on the distribution of the projection angles. In fact, in some situations in cryo-EM, certain orientations are more favorable than others and this is where the assumptions made by other algorithms would fail. The estimation of orientations is based on sound mathematical concepts which rely on HLCC at the foundation. The next step, the alternating gradient descent also does not make any assumption and descends onto the correct structure despite it being a non-convex optimization problem. This is due to an initial estimate provided to us by the moment based approach. 
\noindent
\\
\textbf{Future work:}
There are two major directions for future work.
\begin{itemize}
\item There is an interesting insight that we have observed through these experiments: Despite the promising results of compressed sensing in general image reconstruction from compressive measurements \cite{Candes2008} including in tomography \cite{Wang2010}, it does not achieve an accurate reconstruction in this case where there is uncertainty in the sensing matrix, i.e. in the projection angles. The sparsity-based optimization framework is also very sensitive to the initial estimate in such scenarios, and often does not converge at all. On the other hand, the FBP algorithm, which is used in the gradient descent approach is quite robust to the initial estimate and gives us very accurate reconstructions, despite the inevitable errors (even if they are small in value) in the estimates of the orientations and the shifts. This is even more surprising considering the fact that the FBP does not exploit the powerful sparsity prior that is a common feature of compressed sensing algorithms. A detailed theoretical or analytical study of this surprising observation is a major avenue for future research. 
\item The other important direction is the validation of our algorithm on actual Cryo-EM datasets, insect tomography datasets, and CT reconstructions with patient motion.
\end{itemize}

\bibliographystyle{ACM-Reference-Format}
\bibliography{Bib}

\end{document}